%Paper: hep-th/9306055
%From: MARTINA@le.infn.it
%Date: Fri, 11 Jun 1993 19:23:24 +0300 (CET-DST)

 \magnification=1200  \medskip
\null \vskip 46pt
 \centerline{{\bf CHERN - SIMONS GAUGE FIELD THEORY  }}\medskip
\centerline{{\bf OF TWO -  DIMENSIONAL FERROMAGNETS} } \vskip 1in
 \centerline{L. Martina, O. K. Pashaev $^{*)}$ and G. Soliani }  \vskip .3in
\centerline{{\it Dipartimento di Fisica dell'Universita' and
Sezione INFN,  73100
Lecce, Italy} }  \centerline{{\it and} } \centerline{$^{*)}$
{\it Joint Institute
for Nuclear Research,141980 Dubna, Russia} } \vfill
  \par \centerline{\bf Abstract}\par \medskip\noindent{\it  A
Chern-Simons gauged  Nonlinear Schr\"odinger  Equation is derived from
the continuous Heisenberg model in 2+1 dimensions. The corresponding
planar magnets can be analyzed whithin the anyon theory. Thus,  we
show that static magnetic vortices correspond to the  self-dual Chern -
Simons solitons  and are described by the Liouville equation.  The related
magnetic  topological  charge is associated with the electric
charge of anyons.
Furthermore, vortex - antivortex configurations are described by the
sinh-Gordon equation and its conformally invariant extension. Physical
consequences of these results are discussed. } \vskip 6cm
\hrule width5cm
\medskip \noindent Lecce, June 1993 \eject \baselineskip 20pt \par
 The gauged Chern - Simons models have  stimulated a great attention as
field theories of anyons ( particles whit arbitrary spin and
statistics ) in
(2+1)-dimensional physics [1]. These models are connected with the braid
and quantum group representation theory [2, 3], and  are  applied in
SET TERM/WIDTH=132/PAGE=40
the   treatment of the high-${\rm T_{c}}$ superconductivity phenomena [1,4]
and of  the  Fractional Quantum Hall Effect (FQHE) [5].   It has been
shown that the Chern - Simons gauged Landau-Ginzburg model  plays the
role of effective theory  for the FQHE [6]. In the static limit, the self-dual
version of the  model  possesses soliton solutions [7], which correspond to
Laughlin's quasi-particles  and give a realization  of the anyons.
 On the other hand, the authors have shown recently
that the self-dual Chern-Simons model can be associated with the stationary
2-dimensional continuous classical  Heisenberg model,  using  the tangent
space representation of the corresponding Landau-Lifshitz equation (LLE)
[8].  This formalism arises as a natural generalization of the gauge
equivalence approach developed for  the 1+1 dimensional Heisenberg model,
which in this way is mapped into the Nonlinear Schr\"odinger Equation
(NLSE) [9]. A similar correspondence [10] can be established between the
Ishimori model and the Davey-Stewartson equation, which are the integrable
extensions of the previous models in (2+1)-dimensions. These mappings are
essentially  given by the chiral current on a  Lie algebra, determined by the
original system, and  such   relations can be established independently from
the Lax representation and  the  integrability properties. Of course, we lose
some  nice properties of these mappings from the viewpoint of the
Inverse Spectral Transform, but they can be useful in treating non integrable
models.  By this approach, in 1-space dimension, the Heisenberg model is
mapped into  a  gauged NLSE with zero electric field. Surprisingly enough,
in  2-space dimensions we observe that the continuous Heisenberg model
and, more generally,  the   nonlinear $\sigma-$models  are described as
gauged   Chern - Simons field theories.\par  We notice that the tangent space
 representation is similar to the ${\bf CP}^{1}$ formulation of the $O(3)$
nonlinear $\sigma$-models [11], but they do not coincide. In fact,  in  both
representations the $SU(2)$ group  diagonalizes the spin
variables, belonging to the coset space  $SU(2)/U(1) $,
and  the local
$U(1)$ invariance  arises.  But in the ${\bf
CP}^{1}$ representation the "matter fields"  are identified with
the complex coordinates of
the group,  while in the tangent space formulation the matter fields are
given in terms of the first order derivatives of the group coordinates. As a
consequence, the Gauss law of the Chern-Simons theory
naturally arises and the $U(1)$ gauge
field can  be  interpreted as  a  "statistical gauge" field.
  Moreover,   the  energy
density contains
 the squared moduli of the matter fields and, as a consequence,
  the magnetic energy  is equivalent to   the "number
of particles"  [12].  \par
In the present paper  the non-stationary (2+1)-dimensional
continuous Heisenberg model  is
represented as a two components  NLSE for  the charged matter
fields $\Psi_{\pm}$, interacting
with a Chern-Simons gauge field. Furthermore, the $\Psi_{\pm}$ fields have to
satisfy a supplementary geomentrical constraint and,  for  self-dual
reductions,  we  recover
the model proposed by Jackiw and Pi [7] . The static  magnetic vortices of
our system  are related to  the Chern-Simons solitons and the topological
charge of the former is related to the electric charge of the latter.  Since
the Chern - Simons solitons  are interpreted as   anyons [5], our approach
allows us to describe the planar magnetic states in terms of anyons.
Therefore,  it   explains  the  analogy in the behavior of  the   planar
magnetic vortices and
 the quantum Hall particles, which was mentioned before by  other
 authors [13-14].
 \par  All the results presented here are formally extended to
the case in which the spin variables belong to the non-compact coset space
$SU(1,1)/U(1)$. This model finds several applications,  being
related to antiferromagnetic  [12]
and disordered systems [15]. \par
 The  model we consider is
described by  the   classical spin vector
${\bf S} = \left({{S}_{1},{S}_{2},{S}_{3}}\right)$, where ${\rm
S}_{j}={S}_{j}\left({
t=x_{0},  {x}_{1}, {x}_{2}}\right)$ are real funtions
satisfying the constraint
${{\rm S}_{3}}^{2}+{\kappa}^{2}\left({{{S}_{1}}^{2}+{{S}_{2}}^{2}}\right) = 1$,
and
  $\kappa^{2}=\pm1$ represents the curvature of the spin phase space.
Then, the spin vector belongs to the sphere ${\it S}^{2}  \simeq SU(2)/U(1)$,
or to the pseudosphere ${\it S}^{1,1} \simeq SU(1,1)/U(1)$, when $\kappa
^{2}=1$ or  $\kappa^{2}=-1$, respectively. The time evolution is given by
the LLE  for the continuous isotropic Heisenberg model,
i.e.  $$\partial_{t}  {\bf S}  = {\bf S
}\times \nabla^{2}{\bf S}.\eqno(1) $$   Equation (1)
can be written in matrix form
$$\partial_{t}{S}={1 \over 2i}\left[{S,{\nabla }^{2}S}\right],\eqno(2)$$
where   $\rm S=\left({\matrix{{S}_{3}&\kappa {\overline{S}}_{+}\cr \kappa
{S}_{+}&-{S}_{3}\cr}}\right),$ with $ {S}_{+} = S_{1} + i S_{2}$ and the bar
meaning the complex conjugation. \par
 By
resorting to a right - invariant local $U(1)$ transformation $g$, belonging to
the $ U(2)$ or to the $U(1,1)$  group, the matrix $S$ can be diagonalized in
the form $S = g\sigma_{3} g^{-1}$. In terms of the entries of the matrix $S$,
$g$ is given by $$g=\pm\;
{\left[{2\left({1+{S}_{3}}\right)}\right]}^{-1/2}
\left({\matrix{1+{S}_{3}&{-\kappa
{\overline{S}}_{+}}\cr {\kappa
}^{}{S}_{+}&1+{S}_{3}\cr}}\right)exp\left({i{\sigma }_{3}\beta }\right),
\eqno(3)$$  where we extract the arbitrary real funtion $\beta$. By
definition, the matrix $g$ satisfies the (pseudo-)unitarity condition $\Gamma
g^{\dagger} \Gamma g = {\bf 1}$, where $\Gamma = diag\,(1, \kappa^{2})$.
\par We can introduce
the chiral current $J_{\mu}$ $ \left( \mu = 0, 1, 2\right)$,  which
  reads $${J}_{\mu
}={g}^{-1}{\partial }_{\mu }g={J}_{\mu}^{(0)} + {J}_{\mu}^{(1)}={i \over
4}{\sigma }_{3}{V}_{\mu }+{J}_{\mu }^{(1)},\eqno(4)$$  $$\eqalign{{V}_{\mu
}=\; i{\kappa }^{2}\left[{\left({{S}_{+}{\partial }_{\mu
}{\overline{S}}_{+}-{\overline{S}}_{+}{\partial }_{\mu
}{S}_{+}}\right)/\left({1+{S}_{3}}\right)}\right]+4{\partial }_{\mu }\beta
,\cr}
\eqno(5)$$  when decomposed into the diagonal and off-diagonal parts. The
off-diagonal matrix  ${J}_{\mu }^{(1)}=\left({\matrix{0&{r}_{\mu }\cr
{q}_{\mu }&0\cr}}\right)$ has the non zero components defined by
$${q}_{\mu }={\kappa  \over 2}{S}_{+}\;exp\left({2i\beta }\right){\partial
}_{\mu }\ell n\left({{S}_{+}/\left({1+{S}_{3}}\right)}\right) \qquad
\left({{r}_{\mu }=-{\kappa }^{2}{\overline{q}}_{\mu } }\right).\eqno(6)$$
Thus, the quantities $(q_{\mu}, r_{\mu}, V_{\mu})\; (\mu = 0,1,2)$ are the
tangent space coordinates for the Heisenberg model (1)-(2).   From Eq. (6)
  we  readily obtain  the relation   $$\eqalign{{r}_{\mu }{q}_{\mu
}=-{\kappa}^{2}{\left|{{q}_{\mu }}\right|}^{2}&=-{1 \over
4}\left[{{\left({{\partial }_{\mu }{S}_{3}}\right)}^{2}+{\kappa
}^{2}{\partial }_{\mu }{S}_{+}{\partial }_{\mu
}{\overline{S}}_{+}}\right], } \eqno(7)$$
  which tells
us that the matter density field $\left|{{q}_{m }}\right|^{2}\; \left( m = 1,
2\right)$ is connected with   the magnetic energy density of the model (1)
along the $x_{m}$ - direction  ( henceforth, we will follow the convention to
reserve
 latin indexes,  running on 1, 2, for the space-like quantities, while the
greek indexes can indifferently take values on 0, 1, 2). A relation
similar to (7) holds in 1+1 dimensions as far as the connection between the
NLSE and the  Heisenberg model
 is concerned [12].  \par
{}From definition
(4),  the chiral current $J_{\mu}$ satifies  the zero curvature condition,
namely $${\partial }_{\mu }{J}_{\nu }-{\partial }_{\nu }{J}_{\mu
}+\left[{{J}_{\mu },{J}_{\nu }}\right]=0 .
\eqno(8)$$  \par By using  the
 decomposition (4)  of $J_{\mu}$  and  with the help of Eqs. (6)
and (7), this condition  provides the following equations:
$${D}_{\mu }{q}_{\nu
}={D}_{\nu }{q}_{\mu } \eqno(9)$$ for the off-diagonal part, and
$$\left[{{D}_{\mu },{D}_{\nu }}\right]=-{i \over 2}\left({{\partial }_{\mu
}{V}_{\nu }-{\partial }_{\nu }{V}_{\mu }}\right)=-2{\kappa
}^{2}\left({{\overline{q}}_{\mu }{q}_{\nu }-{q}_{\mu }{\overline{q}}_{\nu
}}\right) \eqno(10)$$ for the  diagonal part, where  we have introduced
 the covariant derivative ${D}_{\mu }={\partial }_{\mu }-{i \over
2}{V}_{\mu }$. \par On the other hand, with the help of   the evolution
equation (2) we can relate ${J}_{0}^{(1)}$ to the other current components,
we get     $${J}_{0}^{(1)}= {1 \over
2}{V}_{m}{J}_{m}^{(1)}+i\,{\partial }_{m}{J}_{m}^{(1)}{\sigma
}_{3},\eqno(11)$$ where the summation over the repeated indices is
understood. Equivalently, we can express the last equation in terms of the
$q_{m}$'s functions, namely $$q_{0} = i\, D_{m} q_{m}.\eqno(12)$$  Equation
(12) allows us  to  eliminate the function $q_{0}$ from the zero curvature
conditions (9-10), obtaining in such a way the  gauge invariant  evolution
equations   $$\matrix{i{D}_{0}{q}_{m} + {D}_{m }{D}_{\lambda }{q}_{\lambda
}=0 ,\cr \cr \left[{{D}_{0},{D}_{m}}\right]= 2i{\kappa
}^{2}\left({{\overline{q}}_{m }{D}_{\lambda }{q}_{\lambda }+{q}_{m
}{\overline{D}}_{\lambda }{\overline{q}}_{\lambda }}\right),\cr
}\eqno(13)$$ and $$\matrix{{D}_{m }{q}_{n
}={D}_{n }{q}_{m }, \cr \cr \left[{{D}_{m },{D}_{n }}\right]= -2{\kappa
}^{2}\left({{\overline{q}}_{m }{q}_{n }-{q}_{m }{\overline{q}}_{n
}}\right).\cr} \eqno(14)$$  Thus, the original LLE (2) is
tranformed into the system of equations  (13) and (14). \par Now,
let us introduce the new fields  $${\Psi }_{\pm }={{q}_{1}\pm i{q}_{2}
\over 2}  \eqno(15)$$and the pair of covariant derivatives defined by
 $${D}_{\pm }={D}_{1}\pm i{D}_{2} = \partial_{1}\pm i\,\partial_{2} -{i\over
2} \left( V_{1}\pm i V_{2}\right) = \partial_{\pm} - {i\over 2} V_{\pm} .
\eqno(16)$$  Notice that, for the complex
 variables $\eta = x_{1} + i\, x_{2},\;  {\overline{\eta}}
= x_{1} - i\, x_{2}$, we
have   $\partial_{-}= 2\, \partial_{\eta}$ and $\partial_{+}= 2 \,
\partial\,_{\overline{\eta}}$. Then,  the system  (13-14) is written
in the
 form $$\eqalignno{{D}_{+}{\Psi }_{-}-{D}_{-}{\Psi
}_{+}&=0&(17.a)\cr
 \left[{{D}_{+},{D}_{-}}\right]=8{\kappa }^{2}&\left({{\left|{{\Psi
}_{+}}\right|}^{2}- {\left|{{\Psi }_{-}}\right|}^{2}}\right)&(17.b)\cr
i{D}_{0}{\Psi }_{\pm }+{D}_{\pm }{D}_{\mp }&{\Psi }_{\pm
}=0&(17.c\,\pm)\cr
 \left[{{D}_{0},{D}_{\pm }}\right]=8i{\kappa}^{2}&\left(\,{{\overline{\Psi
}}_{\mp }{D}_{\pm }{\Psi }_{\mp }+{\Psi }_{\pm}\overline{{D}_{\mp
}}\,\overline{{\Psi }_{\pm }}}\,\right).&(17.d\, \pm)\cr}$$
 \par We
recognize that the  system (17) has the form of a two-component gauged
NLSE. Indeed,   from   (16) and   (17.b) we  can derive   the
identity ${D}_{\pm}{D}_{\mp}$ = $\left({{D}_{1}}^{2} +{{D}_{2}}^{2}\right)
\pm$ 4 $\kappa^{2} \left({{\left|{{\Psi }_{+}}\right|}^{2}- {\left|{{\Psi
}_{-}}\right|}^{2}}\right).$  Then,  by using it into ($17.c\, \pm$)
 and  redefining  the gauge
potential  by
$${A}_{0}={V}_{0}-8{\kappa }^{2}\left({{\left|{{\Psi
}_{+}}\right|}^{2}+{\left|{{\Psi }_{-}}\right|}^{2}}\right),\qquad {A}_{\mu
}={V}_{\mu }\qquad \left({\mu =1,2}\right),\eqno(18)$$  Eqs.  (17b, c, d)
give the system $$\eqalignno{ &i{D}_{0}{\Psi }_{\pm
}+\left({{{D}_{1}}^{2}+{{D}_{2}}^{2}}\right){\Psi }_{\pm }+8{\kappa
}^{2}{\left|{{\Psi }_{\pm }}\right|}^{2}{\Psi }_{\pm }=0&
(19.a\, \pm)\cr &{\partial }_{2}{A}_{1}-{\partial
}_{1}{A}_{2}=8{\kappa }^{2}\left({{\left|{{\Psi
}_{+}}\right|}^{2}-{\left|{{\Psi }_{-}}\right|}^{2}}\right)&(19.b)\cr
&{\partial }_{0}{A}_{j}-{\partial }_{j}{A}_{0}=8{\kappa
}^{2}i{\varepsilon }_{j\ell }\left[\, {{\overline{\Psi }}_{+}{D}_{\ell }{\Psi
}_{+}-{\Psi }_{+}{\overline{D}}_{\ell }{\overline{\Psi
}}_{+}-{\overline{\Psi }}_{-}{D}_{\ell }{\Psi }_{-}+{\Psi
}_{-}{\overline{D}}_{\ell }{\overline{\Psi }}_{-}}\right] &(19.c)\cr}$$ where
the covariant derivative becomes $D_{\mu} = \partial_{\mu} -{i\over 2}
A_{\mu} \left( \mu = 0, 1, 2\right)$.  Furthermore we have the relation
$$\left(D_{1} + i D_{2}\right) \Psi_{-} =  \left(D_{1} -
i D_{2}\right) \Psi_{+}
,\eqno(20) $$  which   is  simply a different form of Eq. (17.a) and can be
considered as a constraint of geometrical origin on the fields.  System (19)
generalizes the class of NLSE interacting with an abelian Chern-Simons gauge
field introduced in Ref.  [7].
 Equations (19.a $\pm$) describe the time evolution of  two opposite
charged  matter fields $\Psi_{\pm}$   in the gauge field $A_{\mu}$,  Eq.
(19.b) defines the "statistical magnetic" field  according to the Chern-Simons
Gauss law and, finally,  Eq. (19.c) defines the "statistical
electric"  field in
the plane.  In fact,  it is easy to verify that the
 source of these fields is the
matter  current $$\matrix{&{\Im}_{\mu }=\left({\rho
,\vec{\Im}}\right)=\left({{\left|{{\Psi }_{+}}\right|}^{2}-{\left|{{\Psi
}_{-}}\right|}^{2},}\right.\cr &\left.{i\left[{{\Psi
}_{+}{\overline{D}}_{m}{\overline{\Psi }}_{+}-{\overline{\Psi
}}_{+}{D}_{m}{\Psi }_{+}+{\overline{\Psi }}_{-}{D}_{m}{\Psi }_{-}-{\Psi
}_{-}{\overline{D}}_{m}{\overline{\Psi
}}_{-}}\right]}\right),\cr}\eqno(21)$$
 which satisfies the continuity equation
$\partial_{\mu}\,\Im_{\mu} = 0$. Therefore, according to  Eq. (7), the
magnetic energy density is related to  the density of particles (anyons) by
$${\partial }_{m}{S}_{3}{\partial }_{m}{S}_{3}+{\kappa }^{2}{\partial
}_{m}{S}_{+}{\partial }_{m}{\overline{S}}_{+}=8{\kappa }^{2}
\left({{\left|{{\Psi
}_{+}}\right|}^{2}+{\left|{{\Psi }_{-}}\right|}^{2}}\right).\eqno(22)$$ \par
 When we limit
ourselves to the {\it static configurations} of the Heisenberg model
(1), it reduces to the 2-dimensional $O(3)$ ($ O\left(2, 1\right)$
respectively)  nonlinear $\sigma$-model. Then,  putting  $V_{0} = 0$, the
system (17)  can be solved   by requiring that
$$\eqalignno{{D}_{\pm }{\Psi }_{\mp }=0, && (23.a \, \pm)\cr
 \left[{{D}_{+},{D}_{-}}\right] = \;&8{\kappa }^{2}\left({{\left|{{\Psi
}_{+}}\right|}^{2}- {\left|{{\Psi
}_{-}}\right|}^{2}}\right).&(23.b)\cr  }$$ \par {\it a) }
System (23) reduces to
the self-dual Chern - Simons model [7] when one of the matter  fields, say
$\Psi_{-}$,
 vanishes.  In this case Eq. (23.a +) is solved by the  gauge
fields $V_{m} $   given by $${V}_{1}=-
{\partial }_{2}\phi  +i{\partial }_{1}\chi , \qquad {V}_{2}= {\partial
}_{1}\phi +i{\partial }_{2}\chi , \eqno(24)$$ where
 $\phi =\ell
n\,{\left|{\Psi_{+} }\right|}^{2}$,  $\chi =\ell
n\left({\overline{\Psi_{+} }/\Psi_{+} }\right)$ . Then, Eq.  (23.b) reduces to
the Liouville equation $$\left({{{\partial }_{1}}^{2}+{{\partial
}_{2}}^{2}}\right)\phi =-8{\kappa
}^{2}e^{\phi}.\eqno(25)$$    This
  equation is well known  in differential geometry and in string theory [16,
17].  Its relation with the ferromagnetic models was recently obtained
 in Ref.  [8]. Here we recall only that it is an integrable equation,
admitting a Lax pair representation.
Moreover,  this equation is invariant under generic conformal
transformations $$\eta = f\left(\tilde{\eta}\right), \overline{\eta} =
g\left(\tilde{\overline{\eta}}\right), \tilde{\phi}\left(
\tilde{\eta},\tilde{\overline{\eta}}\right) =
\phi\left(f\left(\tilde{\eta}\right),  g\left(\tilde{\overline{\eta}}\right)
\right) +  \ell n \left( f'\, g'\right), \eqno(26)$$ where  $f'$, $g'$ are the
derivatives of the functions $f$ and $g$ with respect to their argument. The
general solution of Eq. (25) has the form [7, 18]
$${\left|{\Psi_{+} }\right|}^{2}
= e^{\phi} =
 {{\left|{\partial_{\eta }{\zeta }}\right|}^{2}/
\left({1+{\kappa }^{2}{\left|{\zeta
}\right|}^{2}}\right)^{2}},\eqno(27)$$ where  the
 holomorphic ( except,
maybe, in a finite number of poles)  function $\zeta$
can be reinterpreted as the
stereographic projection $$\zeta= S_{+}/\left( 1 +
S_{3}\right)\eqno(28)$$  of ${\bf S}$ in
the complex plane, as one can realize by replacing Eq. (6) into (15).   The
N-soliton  solutions to Eq (25) is obtained by taking   $\zeta \left({\eta
}\right)=\sum\nolimits\limits_{n=1}^{N} \left[{{c}_{n}/\left({\eta -{\eta
}_{n}}\right)}\right]$, where $c_{n}$ and $\eta_{n}$ are arbitrary complex
numbers.  The  superimposed N-soliton solution at the origin {\bf O}  is
obtained by choosing $\zeta  = a\, \eta^{N}. $
Then, the solitons of the self-dual Chern-Simons  model correspond to the
magnetic vortices (magnetic bubbles) for the static continuous Heisenberg
model.  An  anti-holomorphic stereographic projection
($ \partial_{\, \eta} \zeta = 0$), corresponding  to a   magnetic anti-vortex
configuration,  can be also used in Eq. (27).   Finally, by using the
parametrization (28) for $\bf S$, we obtain a remarkable expression for the
topological charge of this type of solutions, i.e. $$\rm Q=\left({1\over{4\pi
}}\right)\int\!\!\!\int_{}^{}{\bf S} \rm \cdot \left({{{\partial }_{1}{\bf
S}}^{}\rm \times {\partial }_{2}{\bf S}}\right)\rm d{x}_{1}d{x}_{2} = \rm \pm
 {1\over{\pi }} \int\!\!\!\int_{}^{}{\left|{\Psi_{+}
}\right|}^{2}d{x}_{1}d{x}_{2},\eqno(29) $$  (where    the sign -
  is taken when $\zeta$ is anti-holomorphic) which gives  the  charge
quantization for the anyons. In fact,   if the spin field takes values on the
sphere,  the homotopic theorem $\pi\left(S^{2}\right)= Z$  ensures  us that
$Q$ is an integer. An analogous result may be achieved in the non-compact
case.  Relation (29) establishes an explicit  correspondence between the
magnetic topological charge and the electric charge associated with  the
particles of the self-dual Chern-Simons model.

 \par { \it b)} Now let us consider the more general  case, in which neither
$\Psi_{+}$ nor $\Psi_{-}$  is identically zero on the plane. First, we observe
that Eq. (23.a $\pm$) can be written in the form     $${V}_{1}\pm
i{V}_{2}=-2i \partial _{\pm}\ell n\left({\Psi }_{\mp }\right). \eqno(30
\pm)$$ Just by  comparing these equations, we readily obtain that  $$
\partial_{\,\overline{\eta}}\, \ell n \left( \Psi_{-}
{\overline{\Psi}}_{+}\right)
= 0, \eqno(31)$$  then the function
$$U = \ell n \left( \Psi_{-} \overline{\Psi}_{+}\right) \eqno (32)$$ is an
holomorphic function  and it is connected with the conformal symmetry of
the static system.  On the other hand, the existence of  such a function $U$
must  be
compatible with the static Heisenberg model (1). In fact, resorting again to
the expressions of $\Psi_{\pm}$ in terms of the spin variables,  we obtain
$$U=\ell n\left[{{{\partial }_{\eta }\zeta \;
{\partial }_{\eta }\overline{\zeta }
\over {\left({1+{\kappa }^{2}{\left|{\zeta }\right|}^{2}}\right)}^{2}}}\right],
\eqno(33)$$ where the stereographic parametrization (28)  has been used.
Handling this quantity by the help of  the  static limit  of the equation of
motion (1) (or (2)),  we verify  that Eq. (31)   is satisfied.
 \par  Assuming now that $U$ is a
known function, we can solve  Eq.  (32) in the form  $$\Psi_{-} =
{e^{U}\over  \overline{\Psi}_{+}}. \eqno(34)$$
Since the  l.h.s.  of Eq. (23.b)
can be written as $\left[{{D}_{+},{D}_{-}}\right] =
\partial_{-}\,\partial_{+}\, \ell n \Psi_{-} -
\partial_{+}\,\partial_{-}\, \ell n
\Psi_{+}$, we  use the constraint (34) to derive  an equation for
 $\sigma = \ell n \left(\left|{\Psi_{+}}\right|^{2}/
\left|e^{U}\right|\right)$. Indeed   we obtain the
system $$\matrix{\partial_{-} \partial_{+} \, \sigma = &-8\kappa^{J2 }  e^{ F/
2} \left(e^{\sigma} - e^{-\sigma}\right)\cr \partial_{-} \partial_{+} \,F =
&0\qquad\qquad\qquad\qquad\;,\cr}\eqno( 35)$$ where $F = {\rm Re}\,
U$.  First, we observe that  system (35)  is invariant under conformal
transformations   given by
  $$\matrix{ \eta =
f\left(\tilde{\eta}\right) ,&
 \overline{\eta} = g\left(\tilde{\overline{\eta}}\right),&&\cr\cr
\tilde{\sigma} = \sigma\left(f\left(\tilde{\eta}\right),
g\left(\tilde{\overline{\eta}}\right) \right), &\tilde{F} =
F\left(f\left(\tilde{\eta}\right),  g\left(\tilde{\overline{\eta}}\right)
\right) +  2 \,\ell n \left( f'\, g'\right).&\cr} \eqno(36)$$  Secondly, system
(35) is strictly connected with the integrable affine Liouville system
introduced in Ref. [19]. This
 model contains  three fields, one of which, however, is
completely determined by the  functions $\sigma$ and $F$ satisfying
system  (35).  \par By performing a suitable conformal transformation, we
can put $F$ to a constant, then system (35) is reduced to the form
$$\partial_{-}\partial_{+}\, \sigma + \kappa^{2}\, {\rm sinh}\sigma = 0,
\eqno(37)$$ which is the well-known  sinh-Gordon equation. In the context
of the anyon problem it appears as a  special reduction of the SU(2) affine
Toda field theory  [7]. Therefore, it was studied by several authors in
connection with a two-component
 Coulomb gas problem [20, 21]. Equation  (37) is obtained from the Poisson
equation, assuming that
the concentrations of positive- and negative-charged particles
 obey the Boltzmann distribution law $n_{\pm} =  n_{0} \;{\rm exp}\left(\mp
\,\sigma\right)$, where $\sigma$ is the electrical potential.
Then, from this point
of view, system (35) could describe a two-component Coulomb gas with a
non-uniform equilibrium  density   $e^{F/2}$, because of   an external
potential $F$ ( for instance gravitational) satisfying
the Laplace equation. By
analogy with these specific applications, we
can interprete Eqs.(23) as a static
anyon-antianyon system (corresponding to a magnetic vortex-antivortex
system)  at the thermal equilibrium accordingly with the Boltzmann
distibution, with an effective Coulomb interaction.  More generally, in the
non-static case the geometrical constraint (20) my be connected with a
deviation from the thermal equilibrium. \par  The sinh-Gordon equation
arises also in the study of an infinite row of counter-rotating vortices in an
inviscid and incompressible 2-dimensional fluid [22]. Finally,  it appears in
statistical mechanical models of   two-dimensional vortices  systems in a
bounded  magnetized plasma  [23, 24].  The essential point in  the theory of
the sinh-Gordon equation is its complete integrability  [25, 26],  like in the
case of the sine-Gordon equation [27] (which can be obtained from (37)  by
performing the transformation $\sigma \rightarrow   i \, \sigma$) .  Also the
B\"acklund transformations and infinite integrals of motion in
involution can be found  exploiting this mapping between the equations.
 \par  Several
solutions to  Eq. (37) are well known and their main features are connected
with the interaction  signature.  In the case $\kappa^{2} = -1$, cylindrically
symmetric solutions, with  a pointlike  singularity, are well studied, also
from the viewpoint of the Inverse Spectral  Transform  [26]. The asymptotic
behavior at large distances $r \rightarrow \infty$  is given by $\sigma
\rightarrow A\,K_{0}\left(r\right)$, where $K_{0}$ is the modified Bessel
function of the  second kind  of  order 0,  and $A$ is an arbitrary real
constant. However, such a  solution cannot be given in a closed form and it is
related to a Painlev\'e transcendent. There are indications that the
multicharged solutions, i.e.  those solutions whose behavior at infinity is a
linear superposition of a number of cylindrical symmetric singular ones (
one-charge case) centered in different singular points, may exist. However,
this type of  solution  is  no longer   differentiable along some curves
connecting the singular points [28]. \par In the case with  $\kappa^{2} =  1$,
it is possible to construct real analytic solutions [25],
satisfying the condition
$\sigma = 0$ on a rectangular  boundary. Such a solution is given in terms
of the  Riemann theta-function and is parametrized by a set of constants,
called the main spectra, whose interpretation lies in the theory of IST for
periodic boundary condition [29]. A special doubly-periodic solution can be
obtained by separation of variables in the case of square boundary (the
fundamental cell). It reads  $$\sigma = 4\,  {\rm tanh}^{-1} \left[
X\left(x_{1}\right)  Y\left(x_{2}\right)\right], \eqno (38)$$ where
$X\left(x_{1}\right) = k\,  {\rm cn}\left( \overline{x_{1}},
\overline{k}\right) $, $Y\left(x_{2}\right) = {\rm cn}\left( \overline{x_{2}},
\overline{k}\right) $ ,  $ \overline{x_{i}} = x_{i}\, \sqrt{{\left( 1 +
k^{2}\right)/{2 \left( 1 - k^{2}\right)}}}$  $( i = 1, 2)$,  $\overline{k} =
{k/{\sqrt{1 + k^{2}}}}$ and $k$ is determined, for a squared fundamental cell
with a side of unitary length, through the relation $4\,
K\left(\,\overline{k}\,\right) \sqrt{2 \left( 1 - k^{2}\right)/\left( 1 +
k^{2}\right)} = 1 $,  being  $K$ the complete elliptic  integral of  the first
kind.  This solution contains a system of  two vortices and two anti-vortices,
with the nodal lines bisecting the domain.  In other words, this solution
corresponds to a static periodic arrangement of anyons and anti-anyons,
which can be interpreted as a condensed state of such particles [30]. \par
To conclude, in this paper we have shown that the  2-space dimensional
classical continuous Heisenberg model is mapped, via the tangent space
representation,  into a two-component Chern-Simons gauged NLSEs (19),
supplemented by the geometrical constraint (20). Then, according to the
interpretation given by several authors about the role played by  the
Chern-Simons electrodynamics, we can describe the planar magnetic
systems, and more specifically their vortex-antivortex excitations, in terms of
anyons. It is  a remarkable fact that the mathematical
procedure leads directly to the Chern-Simons model from the original
problem, without any further guess concerning the interaction existing
among the magnetic vortices or, alternatively, between the anyons. The
equivalence between magnetic vortices and anyons is stressed in the static
case, where the derived system reduces to
 self-dual Chern-Simons models, leading to the Liouville equation and,  more
generally,  to the sinh-Gordon equation. Both these equations are completely
integrable, as  the original static problem is.
Solutions of these equations are
well-known, and  their interpretation in terms of magnetic vortices is quite
simple. In particular soliton solutions of the Liouville equation correspond to
spin field configurations having analytic stereographic projection. Therefore,
the magnetic topological charge gives the electrical charge of the anyons.
Then, quantization of the electrical charge of anyons has a topological
 explanation.  In the case of the sinh-Gordon equation we have a
two-component system of anyons and anti-anyons at  thermal equilibrium,
representing an analogous combination of magnetic vortices and anti-vortices.
 We have shown that the self-dual system (23) is equivalent to the
Poisson-Boltzmann equation for such a  system. In particular, static
periodic configurations can be obtained for the ordinary $SU(2)$
spin field. For
the non-compact $SU(1,1)$ spin model,
a rich structure of singularities emerges,
but their meaning  is far from being  understood.  Furthermore, by
performing a  conformal transformation on the sinh-Gordon, we can
obtain solutions for the more general system (35).  Finally, in
deriving this
system we have found the quantity U, defined in (33), which
expresses  the
conformal invariance property of the original static Heisenberg model.
 All these results give a more profound  understanding of
the physics of the
anyons and of the magnetic vortices. For instance,  we can  interprete the
skew deflection of magnetic bubbles in a magnetic field gradient [31]
and, in general, the analogies exhibited by magnetic systems and planar
superconductors. \par This work was supported in part by MURST
of Italy and by INFN - Sezione di Lecce. One of the authors
(O. K. P.) thanks
the Department of
 Physics of Lecce University for the warm hospitality. \vfill
\eject \par  {\bf References}\par \medskip\medskip \noindent
\item{[1]} F.
Wilczek, {\it
 Fractional Statistics and Anyon Superconductivity}, World
Scientific, Singapore ( 1990).   \par\medskip \noindent
\item{[2]} E. Witten,
Comm. Math. Phys, {\bf
 120}, 351 (1989). \item{[3]} J. Fr\"olich, in Carg\'ese
Lectures 1987. \item{[4]}A. P. Balachandran et al., {\it
 Hubbard Model and
Anyon Superconductivity}, Lectures Notes in Physycs - Vol. 38, World
Scientific, Singapore ( 1990).  \par \medskip \noindent  \item{[5]}
S. Girvin
and R.
Prange (eds.), {\it The Quantum Hall Effect}, Springer Verlag, New
York (1990). \par\medskip \noindent \item{[6]} S. M. Girvin and A. H.
McDonald, Phys. Rev. Lett. {\bf 58}, 1252 (1987).  \par\medskip \noindent
\item{[7]} R. Jackiw and S. Y. Pi, Progr. Theor. Phys. Suppl. {\bf 107}, 1
(1992).    \par\medskip \noindent \item{[8]} L. Martina, O. K. Pashaev and G.
Soliani, preprint Lecce (1993), submitted to  Journal of Physics A.
\par\medskip \noindent \item{[9]} V. E. Zakharov and L. A. M. Takhtajan,
Theor. Math. Phys{\bf 38}, 26 (1979). \par\medskip \noindent \item{[10]} R.
A. Leo, L. Martina and G. Soliani, J. Math. Phys. {\bf 33}, 1515 (1992).
\par\medskip \noindent \item{[11]} H. Eichenherr, Nucl. Phys. {bf B146},
215 (1978). \par\medskip \noindent \item{[12]}  V. G. Makhankov and O.K.
Pashaev, {\it Integrable Pseudospin Models in Condensed Matter}, in Sov. Sci.
Rev., Math. Phys. {\bf 9}, 1  (1992).\par\medskip \noindent \item{[13]} N.
Papanicolau and T. N. Tomaras, Nucl. Phys {\bf B 360}, 425 (1991).
\par\medskip \noindent \item{[14]} E. G. Floratos, Phys. Lett. {\bf B 279},
117 (1992). \par\medskip \noindent \item{[15]} A. M. M. Pruisken , Nucl.
Phys. {\bf B 290}, 61 (1987).\par\medskip \noindent \item{[16]} J. L.
Gervais, A. Neveu, Nucl. Phys. {\bf B 238}, 125 (1984). \par\medskip
\noindent \item{[17]} L. D. Faddeev, L. A. Takhtajan, {\it Springer Lecture
Notes in Physics}  {\bf 246}, 166 (1986). \par\medskip \noindent
\item{[18]} S.K. Kim, K. S. Soh and J.H. Yee, {\it Phys.
Rev. } {\bf D 42}, 4139 (1990). \par\medskip \noindent \item{[19]} O.
Babelon and L. Bonora, Phys. Lett. {\bf B 244}, 220 (1990). \par\medskip
\noindent \item{[20]} N. Martinov and D. Ouroushev, J. Phys. A: Math. Gen
{\bf 19}, 2707 (1986).  \par\medskip \noindent \item{[21]}   N. Martinov
and N. Vitanov, J. Phys. A: Math. Gen {\bf 25}, L51 (1992). \par\medskip
\noindent \item{[22]} R. Mallier and S. A. Maslowa, Phys. Fluids {\bf A 5},
1074 (1993).  \par\medskip \noindent \item{[23]} D. L. Book {\it et al. },
Phys. Rev. Lett. {\bf 34}, 4 (1975). \par\medskip \noindent \item{[24]} Y. S.
Pointin and T. S. Lundgren, Phys. Fluids {\bf 19}, 1459 (1976).
\par\medskip \noindent \item{[25]} A. C. Ting, H. H. Chen and Y. C. Lee,
Physica {\bf 26 D}, 37 (1987).  \par\medskip \noindent \item{[26]} M.
Jaworski and D. Kaup, Inverse Problems {\bf 6}, 543 (1990). \par\medskip
\noindent  \item{[27]} R. K. Dodd, J. C.
Eilbeck, J. D. Gibbon and H. C. Morris, {\it Solitons and Nonlinear
Wave Equations} (Academic Press, London 1984)\item{[28]}   M. Jaworski
and D. Kaup, in {\it Nonlinear Evolution Equations and Dynamical Systems}, M.
Boiti {\it et al.} Eds.,   Singapore, World Scientific, pag. 118 (1992).
\par\medskip \noindent \item{[29]} B. A. Dubrovin, V. B. Mateev and S. P.
Novikov, Russ. Math. Surv. {\bf 31}, 59 (1976). \par\medskip \noindent
\item{[30]} P. Olesen, Phys. Lett. {\bf B 265}, 361 (1991).\par\medskip
\noindent \item{[31]} T. H. O'Dell, {\it Ferromagnetodynamics, the dynamics
of magnetic bubbles, domains and domain wall}', New York, Wiley (1981).
 \end